# Chip-Level Electromigration Reliability Evaluation with Multiple On-Die Variation Effects

Karthik Airani, Rohit Guttal

Department of Silicon Engineering, Micron Technology

8000 South Federal Way, Boise, Idaho

## Abstract

In this paper, we briefly introduce physical foundations of electromigration (EM) and present a few classical EM-related theories. We discuss physical parameters affecting EM wire lifetime and we introduce some background related to the existing EM physical simulators. In our work, for EM physical simulation we adopt the atomic concentration balance-based model. We discuss the simulation setup and results. We present a variation-aware electromigration (EM) analysis tool for power grid wires. The tool considers process variations caused by the chemical-mechanical polishing (CMP) and edge placement error (EPE). It uses a compact model that features critical region extraction and variation coefficient calculation.

## 1. Introduction

Electromigration (EM) is a phenomenon of material transport caused by a gradual movement of ions in a conductor due to momentum transfer between conducting electrons and diffusing metal atoms [1]. A void or an extrusion may be formed if a wire undergoes EM stress for a sufficiently long period of time. Such defects may eventually damage the wire. For wires manufactured in Dual Damascene process, void formation occurs much faster than extrusion growth. Thus wire damage caused by voids is much more probable than by extrusions and is considered to be the main EM failure mechanism as shown in Figure 1.

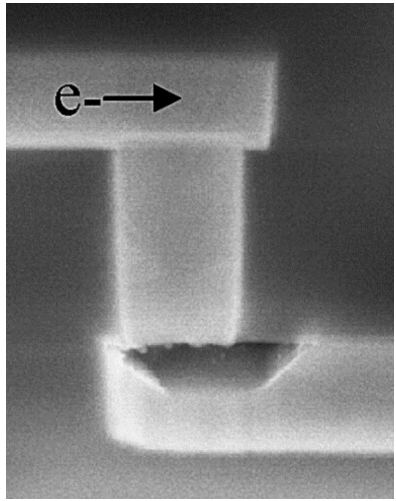

**Fig. 1 EM failures**

In reality, void formation occurs much faster than extrusion growth, which is very unlikely to occur [2]. Therefore voids are considered the major EM-caused failures. In this paper, we only focus on the EM-induced void damage of interconnects.

The ion movement is caused by several forces, the two main ones are: 1) The direct electrostatic force $F_e$ caused by the electric field, and 2) The force from the momentum transfer with moving electrons $F_p$, which is in the opposite direction to the electric field [5]. The resulting force $F_{res}$ acting on an ion is given by (1).

$$F_{res} = F_e - F_p = q(Z_e + Z_p)E = qZ^*j\rho \tag{1}$$

In (1), $q$ is the electron charge, $Z_e$ is the nominal valence of the metal, $Z_p$ accounts for the magnitude and direction of the momentum exchange between the conducting electrons and the metal ions, $Z^*$ is the effective valence, $E$ is the electrical field, $j$ is current density and $\rho$ is the metal resistivity.

## 2. Physics Behind EM
### 2.1 Black's Equation
EM effects have been widely studied. In 1969, J. R. Black developed the equation (referred to as Black's Equation [3]), which captures the EM failure rate dependence on temperature, electrical stress, and material properties.

$$MTTF = Aj^{-n}e^{\left(\frac{E_a}{kT}\right)} \tag{2}$$

In (2), $MTTF$ is the mean-time-to-failure, $A$ is an experimental constant, $j$ is the current density, $n$ is the current density exponent ($1\leq n \leq 2$), $E_a$ is the activation energy, $k$ is Boltzmann constant, $T$ is the absolute temperature in K. Black's equation indicates that larger current density results in shorter EM lifetime. The current exponent $n$ determines whether the shape of a current waveform affects EM lifetime [4]. For $n=1$, the time-average current density can be used in (1) and the current waveform shape does not affect EM lifetime; else if $n>1$, transient current density should be considered and the shape of a current waveform matters. There is some controversy across EM literature regarding the exact value of $n$. A common agreement is that the value of $n$ is close to 1 when the current density is small, and it gradually increases to 2 as the current density increases. For example, in [4], the authors report $n=1$ for $j\leq 0.1$MA/cm$^2$, $n=1.5$ for $0.1$MA/cm$^2<j<1$MA/cm$^2$ and $n=2$ for $j\geq 1$MA/cm$^2$; in [5], the authors report $n=1.1\pm 0.2$ for $j\leq 2.5$MA/cm$^2$ and $n=1.8$ for $j>2.5$MA/cm$^2$. In our experiments, we use the time-average current density on low $j$ wires and triangular waveforms on high $j$ wires. With increasing current densities and decreasing wire dimensions, it is possible that more wires will have current density in the range corresponding to $n>1$. In such cases, more detailed information about current waveforms will be required for EM analysis and discussion on this topic is beyond the scope of this paper.

The model is abstract, not based on a specific physical model, but it flexibly describes the failure rate dependence on the temperature, the electrical stress, and the specific technology and materials. The value of Black's equation is that it maps experimental data taken at elevated temperature and stress levels in short periods of time to the expected component failure rates under actual operating conditions.

### 2.1 Blech Length Effect
For electromigration to occur, there is also a lower limit on the interconnect length [6]. It is known as "Blech length" [7], and any wire that has a length below this limit (typically on the order of 10-100 μm) will not fail by electromigration. Here, the mechanical stress buildup causes a reversed migration process which reduces or even compensates the effective material flow towards the anode (Figure 2).

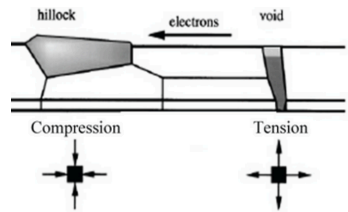

**Fig. 2 Back stress buildup.**

Although known as "Blech length" effect, the actual criterion to determine if a wire is EM-immortal or not, is the product of its current density $j$ and length $L$. Specifically, a conductor line is not susceptible to electromigration if the product of the wire's current density $j$ and length $L$ is smaller than a process technology-dependent threshold value $jL_{threshold}$. Below, we briefly explain the phenomena behind the Blech length effect. The electromigration induced drift velocity is determined by (3).

$$V_d = \frac{D_a |Z^*| e\rho j}{kT} \quad (3)$$

I.A. Blech observed that the metal ions stopped moving, when the wire length was reduced to a certain length. Also, he observed that no ion drift could be detected below a threshold current density.

These observations can be explained by considering the flux due to electromigration and the gradient of the chemical potential via a gradient of mechanical stress [7-10] according to (4)

$$J_v = \frac{D_v C_v}{kT} \left( |Z^*| e\rho j - \Omega \frac{\partial \sigma}{\partial x} \right) \quad (4)$$

where $\Omega$ is the atomic volume, and $\sigma$ is the hydrostatic stress. This equation shows that a gradient of mechanical stress acts as driving force against electromigration. Thus, electromigration stops, when the opposing stress gradient, commonly referred to as "back stress" [11], equals the electromigration driving force, so that $J_v=0$. This steady-state condition is the so-called "Blech Condition", given by (5)

$$\frac{\partial \sigma}{\partial x} = \frac{|Z^*| e\rho j}{\Omega} \quad (5)$$

Integrating (5) over the interconnect line length yields (6)

$$\sigma(x) = \sigma_0 + \frac{|Z^*| e\rho j}{\Omega} x \quad (6)$$

where $\sigma_0$ is the stress at $x=0$. This equation shows that the stress varies linearly along the line, when the backflow flux equals the electromigration flux. Given that the maximum stress a conductor line can withstand is $\sigma_{th}$, the critical product for electromigration failure can be stated as (7)

$$(jL)_c = \frac{\Omega(\sigma_{th} - \sigma_0)}{|Z^*| e\rho} \quad (7)$$

This is the so-called "Blech Product". The critical product provides a measure of the interconnect resistance against electromigration failure and several experimental works have reported that the critical product for modern copper interconnects is in the range from 2000 to 10000 A/cm [12-15].

## 2.3 Diffusion Paths

In a homogeneous crystalline structure, because of the uniform lattice structure of the metal ions, there is hardly any momentum transfer between the conducting electrons and the metal ions. However, this symmetry does not exist at the grain boundaries and material interfaces, so here the momentum is transferred much more vigorously [16]. Since in these regions, the metal ions are bonded more weakly than in a regular crystal lattice, once the electron wind has reached certain strength, atoms become separated from the grain boundaries/interfaces and are transported in the direction of the current. Figure 3 shows the EM diffusion paths in a dual-damascene copper interconnect structure, which is the main metallization process in modern semiconductor fabrication [17].

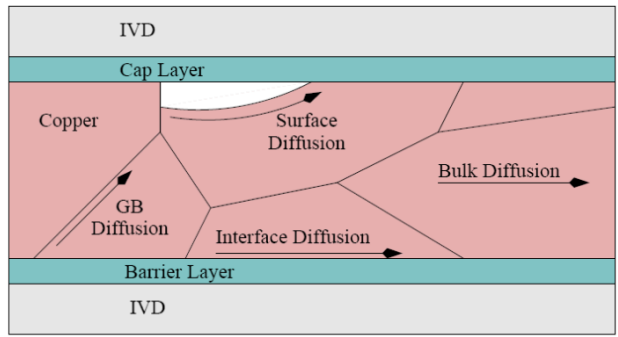

**Fig. 3. Dual-damascene copper EM diffusion paths**

In a dual-damascene structure, only a single metal deposition step is used to simultaneously form the main metal line and the via beneath it. After the via and trench recesses are etched, the via is filled in the same metal-deposition step that fills the trench. Therefore, vias are always in the same copper (Cu) body as the upper level metal, but are separated from the lower level metal by a barrier layer which prevents copper diffusion into the dielectric as shown in Figure 3. Typical barrier layer materials are Ta, TaN, TiN and TiW. The diffusion processes caused by electromigration can be divided into grain boundary diffusion, bulk diffusion and surface diffusion. In general, surface diffusion is dominant in copper interconnects.

## 3. Variation-Aware Compact MTTF Model
### 3.1 Variation Sources
#### 3.1.1 CMP dishing

Chemical Mechanical Polishing/Planarization (CMP) is a process of smoothing surfaces with the combination of chemical and mechanical forces [18]. Copper and the adjacent dielectric are removed from the wafer at different rates during CMP, creating surface anomalies and a varying topography [19]. Many factors, including pattern geometry (e.g., line density), affect the material removal rates. The CMP process strives to achieve flat topography to improve yield. Dishing is a copper surface anomaly caused by CMP. Dishing occurs when copper recedes below the level of the adjacent dielectric. Although dummy metal fills are added in order to achieve a flat topography, dishing still occurs (Figure 4).

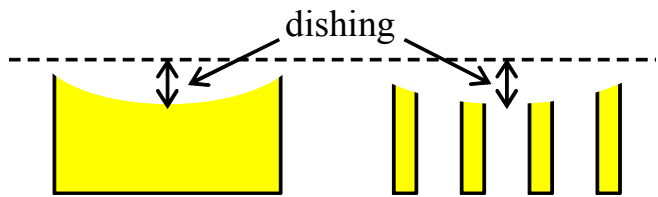

**Fig. 4 CMP dished copper wires.**

#### 3.1.2 Edge Placement Error

As technology continues to scale down, lithography no longer produces ideal geometric shapes. The most typical error is the Edge Placement Error (EPE) [20]. EPEs outside the feature are considered positive errors (bumping) and EPEs inside the feature are negative errors (necking) as shown in Figure 5. Although techniques such optical proximity correction (OPC) are applied to reduce EPE, these variations still exist.

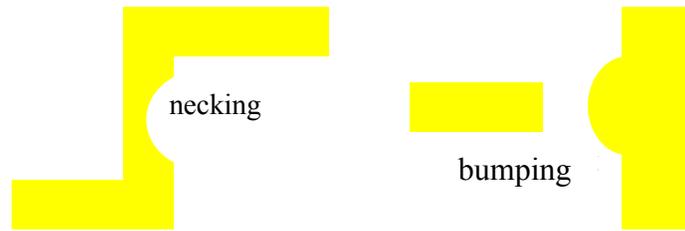

**Fig. 5. EPE affected wires**

## 3.2 Compact Model
### 3.2.1 Observations
Figure 6 shows an example of AFD simulation result by ANSYS of a CMP dished wire. A maximum AFD occurs at the cathode and a minimum AFD at the anode end of the wire.

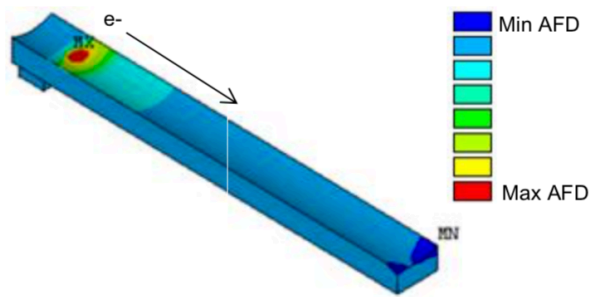

**Fig. 6 EM physical simulation of a CMP dished wire.**

We make several observations based on the simulations with CMP/EPE variations:

1) A bumping on a wire does not relax the overall EM stress, whereas necking reduces the EM lifetime;

2) On a wire, there exists a critical region where maximum AFD occurs. Necking in this region affects EM significantly; necking elsewhere affects EM much less. The position of a critical region depends on the wire length, current density, and current direction;

3) When CMP dishing affects the critical region, the resulting EM lifetime is proportional to the square of the wire height;

4) When necking affects the critical region, the resulting EM lifetime is a function of the necking size and location.

These observations inspire us to develop a compact model that captures dependencies between wire geometry imperfections and EM lifetime. We run simulations with various configurations, and develop a reasonably accurate model within a practical range of geometric variations.

### 3.2.2 Critical Region

We define the critical region of a wire as a region where the *AFD* is greater than 10% of $AFD_{max}$ (the maximum value of *AFD* along the entire wire). The critical region boundary position is found to be a piecewise function of wire length and current density as shown in (8), where $k_1$=0.85, $k_2$=1.65e-12, $C_R$=0.85e6, $C_L$=1.02e16 and $C_C$=0.0915 are constants, $l$ is the wire length and $j$ is the current density.

$$B_L = \begin{cases} 0 & 0 \leq l < \frac{C_L}{k_1 j} \\ k_1 l - \frac{C_L}{j} & l \geq \frac{C_L}{k_1 j} \end{cases}$$

$$B_R = \begin{cases} k_2(j + C_C) \cdot l & 0 \leq l < \frac{C_R}{k_1 j^2} \\ k_1 l - \frac{C_R}{j^2} & l \geq \frac{C_R}{k_1 j^2} \end{cases} \quad (8)$$

A sample curve for current density 40mA/mm² is shown in Figure 7. In this Figure, the square- and circle-marked lines represent distances from the cathode via to the right and left boundaries of the critical region. The double arrowed lines mark the lengths of critical regions. It can be observed that for a fixed current density, the length of critical region does not change.

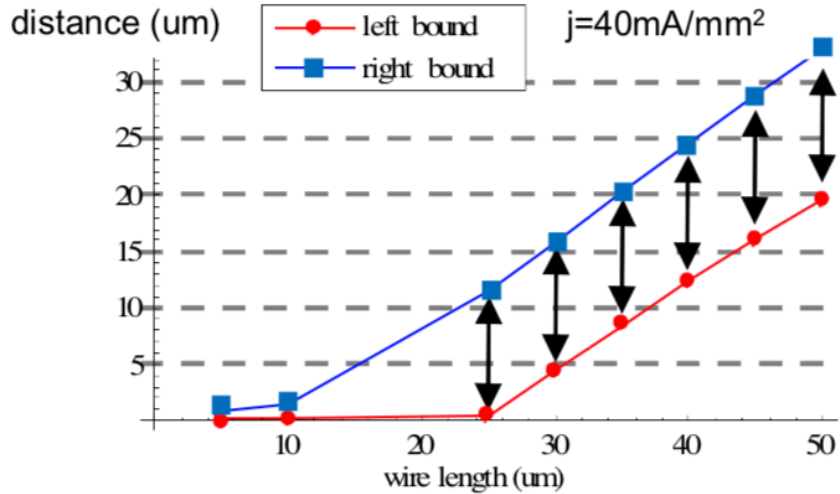

**Fig. 7 Critical region boundaries.**

To quantify the effects of process variation we introduce the variation coefficient.
**Definition 1:** Variation coefficient $C_{var}$ is the ratio of an ideal wire EM lifetime $t_{life0}$ and EM lifetime $t_{life}$ of the corresponding wire with geometric variations.

$$C_{var} = \frac{t_{life0}}{t_{life}} \quad (9)$$

Unlike for process variation induced global effects on EM lifetime, when considering local effects, we assume that the amount of charge passing through a wire does not change.

### 3.2.3 Compact Model

There are many factors that affect the EM lifetime of a wire with CMP/EPE variations. However, instead of investigating those detailed physical effects, we try to extract the major components that are most important to EM lifetime with variations and develop a simplified model that can be easily applied to full-chip interconnect EM analysis.

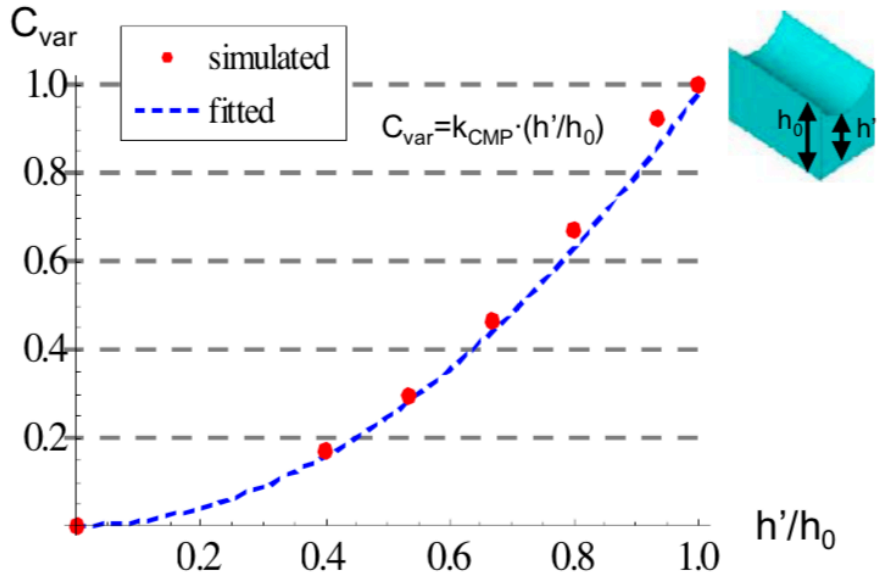

**Fig. 8. Compact model curve for $C_{var}$ due to CMP.**

The ratio between an imperfect and ideal wire height ($h'/h_0$) is the dominant factor for $C_{var}$ of CMP dishing. This dependency is given by (10), and shown in Figure 3.5, for $k_{CMP} = 1$ and $m = 2$.

$$C_{var} = k_{CMP} \left( \frac{h'}{h_0} \right)^m \tag{10}$$

To determine $C_{var}$ of EPE necking, we first consider the case where a single circular dent is present on a wire. The dent can be completely described by its depth $d$, length $l$ and location $x$, and $C_{var}$ is a function of these three parameters captured by (11), where $b$ and $w$ are the length of the critical region and the width of a wire; $k_{EPE}$ is a fitted parameter equal to 7.2e18. Figure 3.6 shows a sample plot for $C_{var}$ of EPE necking with $d=0.03\mu m$ and $l=1\mu m$.

$$C_{var} = f(\frac{d}{w}, \frac{l}{b}, x) = \begin{cases} 1 & x \leq B_L \text{ or } x \geq B_R \\ 1 + k_{EPE}\frac{1}{b}\frac{d}{w}(x - B_L) & B_L < x \leq (B_L + B_R)/2 \\ 1 - k_{EPE}\frac{1}{b}\frac{d}{w}(x - B_R) & (B_L + B_R)/2 < x < B_R \end{cases}$$

(11)

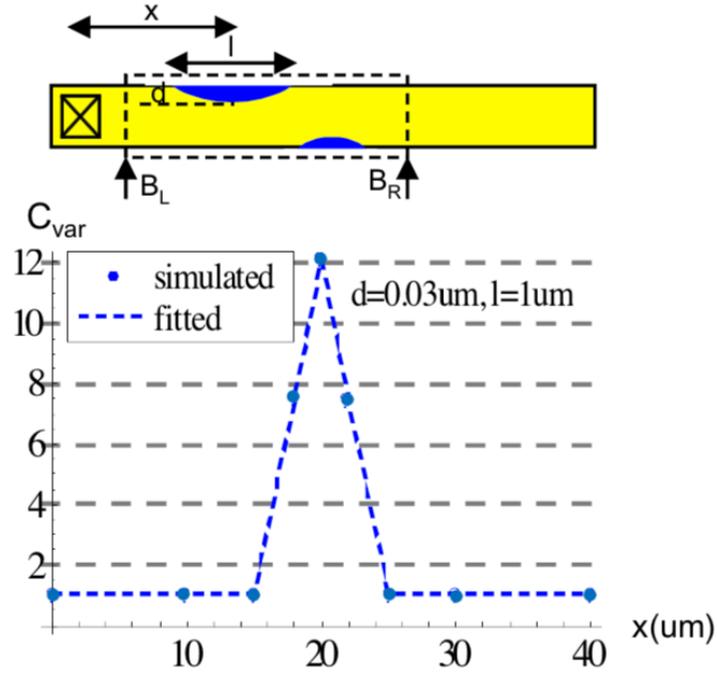

Fig. 9. Compact model curve for $C_{var}$ due EPE

When multiple dents are present on a wire, their cumulative effect is found to be the product of $C_{var}$ caused by each dent, given by (12).

$$C_{var} = \prod_i C_{var,i}$$

(12)

This variation-aware EM lifetime model provides a mapping methodology from an assumed variation to the resulting EM lifetime. A more useful way to apply this model is to estimate the variation tolerance of EM-sensitive wires. For example, we may want to know how much CMP dishing or how much EPE is allowed for a given wire. These are later referred to as variation criticalities, and will be given to designers.

## 6. Conclusions

As technology scales down, EM is becoming progressively a more serious problem. It posts huge challenge on analysis tools and limits the application of conventional EM theories such as Black's equation and Blech length effect. Moreover, the process variation brings in more uncertainty to EM analysis.

In this paper, we introduced the basic background knowledge of EM, and described the modeling of variation effects. The variation effects mainly refer to the changes of wire geometry caused by CMP and EPE that affect the EM lifetime of individual wires. A compact variation-aware EM lifetime model is developed. Using this model, variation tolerance of each EM-sensitive wire can be fed back to designers. As technology is scaling down, process variation on wires becomes increasingly more significant. Our work shows that its effect on EM reliability is non-negligible, and a variation-aware EM analysis tool may provide a more realistic assessment of the EM reliability of power grids.